\documentclass[aps,prb,amsmath,amssymb,footinbib,showpacs,twocolumn,superscriptaddress]{revtex4-1}
\usepackage{amsmath}
\usepackage{amssymb}
\usepackage{amsthm}
\usepackage{setspace}
\usepackage{graphicx}
\usepackage{braket}
\usepackage{mathrsfs}
\usepackage{float}
\usepackage[colorlinks = true,linkcolor = red,urlcolor  = blue,citecolor = blue,anchorcolor = blue]{hyperref}
\usepackage[utf8]{inputenc}
\usepackage[english]{babel}
\usepackage{bm}

\begin{document}

\title{Majorana versus Andreev bound state energy oscillations in a 1D semiconductor-superconductor heterostructure}

\author{Girish Sharma}
\affiliation{School of Basic Sciences, Indian Institute of Technology Mandi, Mandi-175005 (H.P.), India}

\author{Chuanchang Zeng}
\affiliation{Department of Physics and Astronomy, Clemson University, Clemson, South Carolina 29634, USA}

\author{Tudor D. Stanescu}
\affiliation{Department of Physics and Astronomy, West Virginia University, Morgantown, West Virginia 26506, USA}

\author{Sumanta Tewari}
\affiliation{Department of Physics and Astronomy, Clemson University, Clemson, South Carolina 29634, USA}

\begin{abstract}
The recent experimental observations of decaying energy oscillations in semiconductor-superconductor Majorana nanowires is in contrast with the typical expectations based on the presence of Majorana zero modes localized at the ends of the system, when the amplitude of the hybridization energy oscillations is predicted to increase with the applied magnetic field. These observations have been theoretically justified recently by considering a position-dependent,  step-like spin-orbit coupling near end of the nanowire, which could arise due to the presence of tunnel gates in a standard tunneling conductance experiment. Here, we show that the window in parameter space where this phenomenology occurs is vanishingly small, when compared to  the parameter region where Majorana oscillations increase in amplitude with the applied field. Further, including a position-dependent effective potential, which is also induced naturally near the end of the wire by, e.g., tunnel gates, practically removes the small window associated with decaying oscillations. Using extensive numerical calculations, we show that, as expected, increasing amplitude oscillations of the hybridization energy represent a generic property of topological Majorana zero modes, while decreasing amplitude oscillations are a generic property of low-energy trivial Andreev bound states that typically emerge in non-homogeneous systems. By averaging over several realistic parameter configurations, we identify robust features of the hybridization energy that can be observed in a typical differential conductance experiment without  fine-tuning the control parameters.
\end{abstract}

\maketitle

\section{Introduction}

One of the most promising platforms predicted to support emergent non-Abelian Majorana fermions in a solid-state system consists of a strong spin-orbit coupled semiconductor nanowire with proximity induced superconductivity and magnetic field applied parallel to the wire~\cite{sau2010generic, sau2010non,oreg2010helical,lutchyn2010majorana}. The prospect of employing this type of  platform for fault tolerant topological quantum computation (TQC)~\cite{kitaev2001unpaired,nayak2008non} has spurred tremendous experimental progress in recent years~ 
\cite{mourik2012signatures,deng2012anomalous,das2012zero,rokhinson2012fractional,churchill2013superconductor,finck2013anomalous,deng2016majorana,zhang2017ballistic,chen2017experimental,nichele2017scaling,zhang2018quantized,albrecht2017transport,o2018hybridization,shen2018parity,sherman2017normal,vaitiekenas2018selective,albrecht2016exponential}.
When the wire is in the topological phase, two Majorana zero modes (MZMs) -- sometimes called zero-energy Majorana bound states (MBSs) -- are predicted to emerge at the opposite ends of the system, which gives rise to a quantized zero-bias conductance peak of height $2e^2/h$ for charge tunneling into the ends of the wire~\cite{sengupta2001midgap,law2009majorana, flensberg2010tunneling}. 
A major experimental advancement has been the recent observation by Zhang \textit{et al}.~\cite{zhang2018quantized} of a quantized zero-bias conductance peak of height $2e^2/h$ corresponding to finite length quantized conductance plateaus as function of control system parameters such as the Zeeman field and gate potentials. 
In non-ideal heterostructures, other mechanisms, such as disorder~\cite{bagrets2012class,liu2012zero,degottardi2013majorana,degottardi2013majoranaprl,adagideli2014effects,rainis2013towards}, quantum interference~\cite{pikulin2012zero}, system inhomogeneity~\cite{roy2013topologically,san2013multiple,kells2012near,chevallier2012mutation,ojanen2013topological,stanescu2014nonlocality,san2016majorana,klinovaja2015fermionic,fleckenstein2018decaying,cayao2015sns}, and coupling to a quantum dot~\cite{prada2012transport,liu2017andreev}, can also generate near-zero topologically-trivial states that give rise to zero-bias peaks in the $dI/dV$ spectra in the absence of Majorana zero modes. In general, the height of a zero-bias conductance peak associated with a trivial  state is not quantized. If there is an accidental quantization at some specific parameter values, this is not expected to be robust against relatively small changes of the system parameters, in contrast to a ``true'' topological Majorana peak which should be robust against any perturbation that does not drive the system out of the topological phase.

While the emergence of quantized conductance plateaus  is a necessary {\em consequence} associated with the presence of topological Majorana zero modes localized at the opposite ends of the nanowire, recent  theoretical studies~\cite{moore2018two,moore2018quantized,vuik2018reproducing} have shown that the presence of MZMs does not represent a necessary {\em condition} for quantized plateau in local charge tunneling conductance measurements. Indeed, a quantized conductance plateau of height $2e^2/h$ can also emerge in a topologically trivial system due to the presence of Andreev bound states (ABSs) having the component Majorana modes partially separated in space. These ABSs are referred  as the partially separated Andreev bound states (ps-ABSs)~\cite{moore2018quantized,moore2018two}, while the corresponding component Majorana states were dubbed quasi-Majorana modes~\cite{vuik2018reproducing}. These states are topologically trivial,  they are more sensitive to variations of the system parameters than the topological MZMs (although relatively robust, as compared to ``standard'' ABSs consisting of completely overlapping component MBSs),  and cannot be harnessed for topological quantum computation. Distinguishing these doppelganger states from the true topological MZMs is extremely important, both in principle -- to unambiguously demonstrate the realization of genuine MZMs -- and as a practical requirement for building Majorana-based topological qubits. This could be done using interference type measurements~\cite{sau2015proposal}, or, within the current state-of-the-art, looking for correlations between (local) charge tunneling measurements done at both ends of the wire in a three-terminal experiment\cite{moore2018two}. 

Partial information about the spatial separation of a (local) MBS pair can be obtained even from a single local measurement, by analysing the dependence of the energy splitting generated by the partial overlap of the two MBSs on various control parameters.
In a finite length topological nanowire, the Majorana modes located at the two ends of the system have a finite overlap, which leads to the hybridization of the MBSs into finite energy Bogoliubov eigenstates. The corresponding hybridization energy $\delta E$ is predicted to oscillate as a function parameters such as the wire length, magnetic field, and chemical potential~\cite{lin2012zero,cheng2009splitting,sarma2012splitting}. Moreover, based on theoretical modeling, the amplitude of these Majorana oscillations is expected to increase with increasing magnetic field or decreasing wire length. By contrast, in experiment the amplitude of the energy oscillations is seen to either die away after an overshoot or decay with increase in magnetic field field~\cite{albrecht2016exponential,albrecht2017transport,shen2018parity,sherman2017normal,o2018hybridization,vaitiekenas2018selective}. Recently, Cao \textit{et al}.~\cite{cao2019decays} have provided a possible explanation for this observation by considering a position-dependent spin-orbit coupling. More specifically, a step-like profile of the spin-orbit coupling strength near the end of the nanowire (which is physically justified as an effect of applying a tunnel gate potential inside a segment of the wire not covered by the superconductor) is shown to give rise, in certain conditions,  to decaying Majorana oscillations similar to those observed experimentally. 

In this paper, we reexamine the problem of Majorana oscillations by systematically exploring the relevant parameter space and carefully taking into account contributions from various factors associated with the presence of a tunnel gate region. We find that: (i) The window in parameter space associated with decaying energy oscillations of topological MBSs localized near the ends of the wire is vanishingly small, as compared to the region characterized by increasing oscillations. Consequently, observing Majorana oscillations with an amplitude that decreases with increasing magnetic field requires fine-tuning of the chemical potential, magnetic field, and effective potential profile near the end of the wire. (ii) By contrast, decaying energy oscillations associated with (topologically-trivial) quasi-Majorana modes (i.e., ps-ABSs) are rather generic. Consequently, the observation of decaying energy oscillations that persist within a finite parameter space region is a strong indication of ps-ABSs (i.e. quasi-Majoranas). 
Our conclusions are based on model calculations of semiconductor-superconductor hybrid devices that expand the results of the minimal model by incorporating basic experimentally-relevant features.  In a typical charge tunneling experiment, a segment near the end of the nanowire is not proximitized and acts as a tunnel barrier, i.e., it has a small (or vanishing) induced  pairing potential $\Delta$ and is characterized by a position-dependent effective potential profile induced by a tunnel gate. 
This portion of the wire can be modeled  as a quantum dot with a finite potential $V_{dot}$ and an induced gap $\Delta = 0$~\cite{moore2018quantized,moore2018two}. Further, this segment can also have a different value of the spin-orbit coupling (as compared to the spin-orbit coupling in rest of the nanowire) due to the non-uniform gate-induced electrostatic potential.
We parametrize the non-uniformity of the induced pairing by $\eta$ and the spin-orbit field asymmetry by $\lambda$ [see Eqns. (\ref{lambda}) and (\ref{eta})]. When $\eta=0$ ($\lambda=0$), there is no induced pairing (spin-orbit) asymmetry, i.e., the tunnel gate region has the same properties as the rest of the wire (the pristine nanowire limit). On the other hand,  when $\eta=1$ ($\lambda=1$), the induced pairing (spin-orbit coupling) vanishes in the quantum dot region. Note that, typically, one expects $\eta=1$ (i.e., no induced pairing in the tunnel region), but we expand the analysis into the $\eta<1$ regime. We find no significant dependence on this parameter for $\eta \gtrsim 0.75$, i.e., when the induced pairing in the tunnel region is significantly smaller than the pairing in the proximitized wire. 
We first consider the case discussed in Ref.~\onlinecite{cao2019decays} (corresponding to $\eta=0$ and $\lambda>0$) and show that the parameter region where decaying Majorana oscillations are observed is very small compared to the region characterized by Majorana oscillations that increase in amplitude with the applied field. The parameter window characterized by decaying oscillations is restricted to near-zero chemical potential and Zeeman fields not exceeding $1.5h_c$, where $h_c$ is the critical field associated with the topological quantum phase transition.
 Next, we consider a more general scenario, corresponding to $0<\eta<1$, $0<\lambda<1$, and a nonzero quantum dot potential $V_{dot}$, and show that in the presence of a finite tunnel potential the window of decreasing Majorana oscillations practically vanishes. Based on extensive numerical calculations, we show that increasing amplitude oscillations of the hybridization energy $\delta E$ represent a generic property of Majorana modes, while decreasing oscillations are a property of (partially-separated) Andreev bound states. By averaging over several realistic parameter configurations, we identify robust features of the hybridization energy that can be observed in a typical differential conductance experiment without  fine-tuning the control parameters.


\section{Theoretical modeling}~\label{modeling} 

The most common paradigm for engineering a Majorana nanostructure involves a spin-orbit coupled semiconductor nanowire in the presence of a magnetic field (typically oriented along the wire) and proximity-coupled to an $s$-wave superconductor that induces a finite pairing potential.  The simplest effective Bogoliubov-de-Gennes (BdG) Hamiltonian describing the low-energy physics of the ``ideal'' hybrid system can be written as 
\begin{equation}
H_0 = \left(\frac{-\partial_x^2}{2m} -\mu + i\alpha\partial_x \sigma_y\right)\tau_z + h\sigma_z + \Delta \tau_x,       \label{Eq_1}
\end{equation}
where $\sigma_i$ and $\tau_i$ are Pauli matrices associated with the spin and particle-hole degrees of freedom, respectively, $m$ is the effective mass, $\mu$ is the chemical potential, $\alpha$ is the strength of the spin-orbit coupling, $h$ is the applied magnetic (Zeeman) field, and $\Delta$ is the induced  superconducting gap. When solved with open boundary conditions, the Hamiltonian (\ref{Eq_1}) admits zero-energy Majorana modes if the applied magnetic field $h$ exceeds a critical value $h_c = \sqrt{\Delta^2+\mu^2}$ corresponding to the topological quantum phase transition (TQPT) between the trivial and topological superconducting phases. These Majorana modes emerge as bound states localized at the two ends of the wire and must have exactly zero energy in the infinite wire limit (i.e., for $L\gg \xi$, where $\xi$ is the characteristic length scale of the MBS and $L$ the wire length). Given the exciting prospect of realizing Majorana fermions based on this relatively simple setup, the system has been studied extensively in recent years, both theoretically and experimentally. 

In a real, finite system, the Majorana modes localized at the opposite ends of the wire hybridize acquiring a finite energy. The hybridization energy is predicted to oscillate as a function of parameters such as the wire length,  magnetic field, or chemical potential. If large-enough, these energy oscillations result in splitting oscillations of the (near) zero energy conductance peak, which should be experimentally observable. Features consistent with this expectation have been observed in the Coulomb blockade regime of (hybrid) superconducting islands~\cite{albrecht2016exponential,albrecht2017transport,shen2018parity,sherman2017normal,o2018hybridization,vaitiekenas2018selective}. 
However, in contrast to previous theoretical predictions, the amplitude of the observed oscillations decreases with increasing magnetic field. 
To fully understand the significance of these experimental findings, one has to take into account details of the real system that are not included in the simple model given by Eq. (\ref{Eq_1}). For example, in tunneling experiments a small segment near the end of the nanowire is not covered by the superconductor and serves as a tunnel barrier. This segment is characterized by a suppressed (vanishing) induced gap and a finite effective potential controlled by a tunnel gate. One can effectively model this portion of nanowire as a quantum dot with a finite potential $V_{dot}$ and pairing potential $\Delta = 0$. 
The presence of the quantum dot can be incorporated into the model by considering a position-dependent induced gap, $\Delta \rightarrow \Delta(x)$, and adding an effective potential term $V(x)$ that has the value $V_{dot}$ inside the dot region  and is zero otherwise. We note that this step-like potential is the simplest way to model the quantum dot-wire system; the actual effective potential may have a rather complicated profile, as suggested by recent Schrodinger-Poisson calculations~\cite{woods2018effective}.  

\begin{figure}[t]
    \centering
    \includegraphics[scale=0.75]{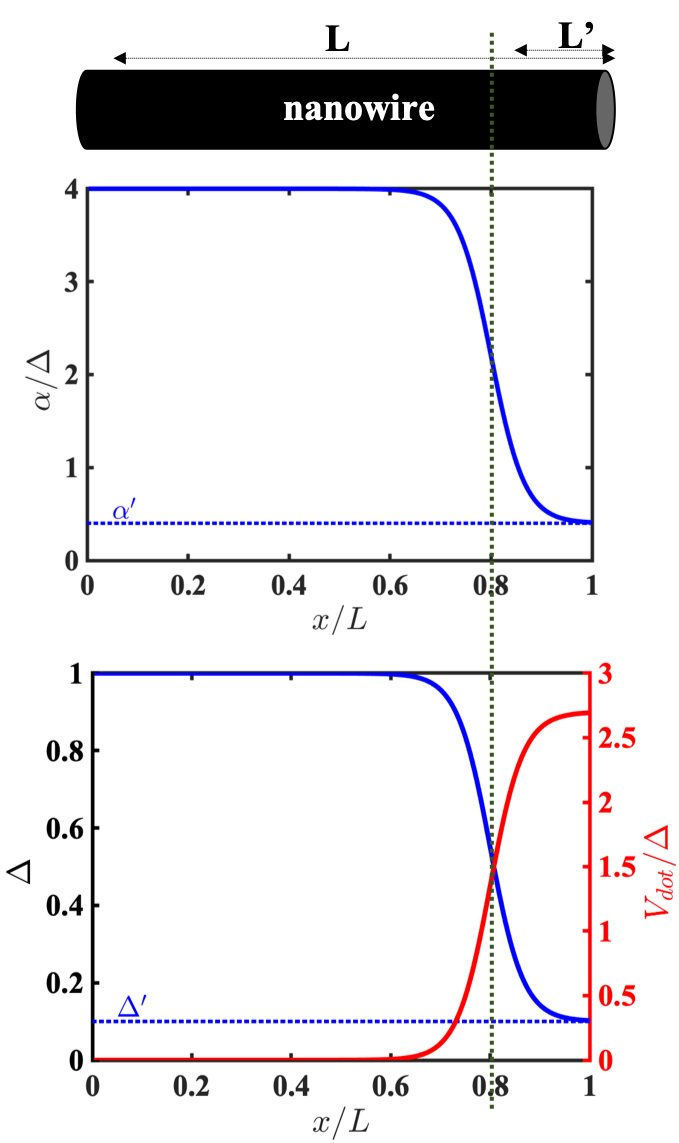}
    \caption{Schematic representation of the spatial profiles of the spin orbit coupling, $\alpha(x)$, induced pairing potential, $\Delta(x)$, and effective potential, $V(x)$, for a Majorana nanowire of length $L$ with an uncovered segment (quantum dot) of length $L^\prime$ at the right end. The dotted line at $x_0=L-L^\prime$ separates the proximitized wire from the quantum dot.} \label{Fig_schematic}
    \vspace{-2mm}
\end{figure}

Recently, it has been pointed out by Cao \textit{et al.}~\cite{cao2019decays} that the spin-orbit coupling term may also have a non-uniform profile in the vicinity of the quantum dot region. This is due to the fact that the effective spin-orbit coupling depends on the component of the electric field perpendicular to the wire, which is manifestly position-dependent in the presence of a potential barrier created by, e.g., a narrow back gate placed under the quantum dot (i.e. the uncovered wire segment). 
This can be incorporated into the model by making the spin-orbit coupling  in Eq. (\ref{Eq_1}) position-dependent, $\alpha \rightarrow \alpha(x)$, where, to a first approximation,  $\alpha(x)$ has a step-like profile.  Using this model, Cao \textit{et al.} predicted~\cite{cao2019decays} the emergence of Majorana oscillations with decaying amplitude, consistent with the recent experimental observations. However, the study does not clarify how generic this property is, i.e., where in the parameter space one can obtain decaying (rather that increasing) energy splitting oscillations as function of the applied field. Moreover, the model used the calculations is fundamentally incomplete, as the presence of the quantum dot potential -- the main reason for having a non-homogeneous spin-orbit coupling -- is neglected.
To better capture the possible consequences of having a non-homogeneous quantum dot-wire system, we generalize the model described by Eq. (\ref{Eq_1}) by including  the gate potential and considering the position-dependence of the effective parameters. Explicitly, we have
\begin{eqnarray}
H &=& \left[\frac{-\partial_x^2}{2m} -\mu + i\alpha(x)\partial_x \sigma_y + V(x)\right]\tau_z  \nonumber\\
   &+&   h\sigma_z + \Delta(x) \tau_x.     \label{Eq_4}
\end{eqnarray}
For simplicity, we assume that the position-dependent parameters have (smoothed) step-like profiles. Specifically, for a wire of total length $L$ having a quantum dot (i.e., an uncovered region) of length $L^\prime$ near its right end, we have 
\begin{eqnarray}
V(x) &=& V_{dot} \frac{1 + \tanh[(x-x_0)\beta_V]}{2}, \\
\alpha(x) &=&  (\alpha-\alpha^\prime)\frac{1 - \tanh[(x-x_0)\beta_\alpha]}{2} + \alpha^\prime, \\
\Delta(x) &=&  (\Delta-\Delta^\prime)\frac{1 - \tanh[(x-x_0)\beta_\Delta]}{2} + \Delta^\prime,
\end{eqnarray} 
where $x_0=L-L^\prime$, $\alpha$ and $\alpha^\prime$ are the spin-orbit coupling strengths in the proximitized wire and dot region, respectively,  wile  $\Delta$ and $\Delta^\prime$ are the corresponding values of the induced gap. The parameters $\beta_V$,  $\beta_\alpha$,  and $\beta_\Delta$ represent the inverse length scales over which the corresponding parameters change from the values corresponding to the proximitized region to the quantum dot values. The spatial profiles of the position-dependent parameters are shown schematically in Fig.~\ref{Fig_schematic}. Note that the dot region corresponds to $x> x_0=L-L^\prime$. 

The parameters  $\lambda$ and $\eta$ that describe the non-homogeneity of the spin-orbit coupling and induced pairing, respectively, are defined as 
\begin{eqnarray}
\lambda &=& 1-\frac{\alpha^\prime}{\alpha}, \label{lambda} \\
\eta &=& 1-\frac{\Delta^\prime}{\Delta}.      \label{eta}
\end{eqnarray}  
As mentioned before, $\eta=0$, $\lambda=0$ corresponds to a homogeneous system (i.e. no quantum dot), while $\eta=1$, $\lambda=1$ corresponds to the vanishing of the effective parameters inside the dot region, i.e., $\alpha^\prime=0$ and $\Delta^\prime=0$. 
In the calculations, we allow the parameters $\eta$ and $\lambda$ to vary independently. 
Finally, we discretize Eq.~\ref{Eq_4} on a 1D lattice and solve the corresponding tight-binding model numerically.
Unless otherwise stated, the following parameter values have been used: $L\approx 3~\mu$m, $L^\prime/L = 0.2$, $\beta_V =\beta_\Delta = \beta_\alpha=0.05/a$, $t=38\Delta$, $\alpha=4a\Delta$, where $a= 10~$nm is the lattice spacing, and $N=300$ lattice sites.

\begin{figure*}[t]
    \centering
    \includegraphics[width=2\columnwidth]{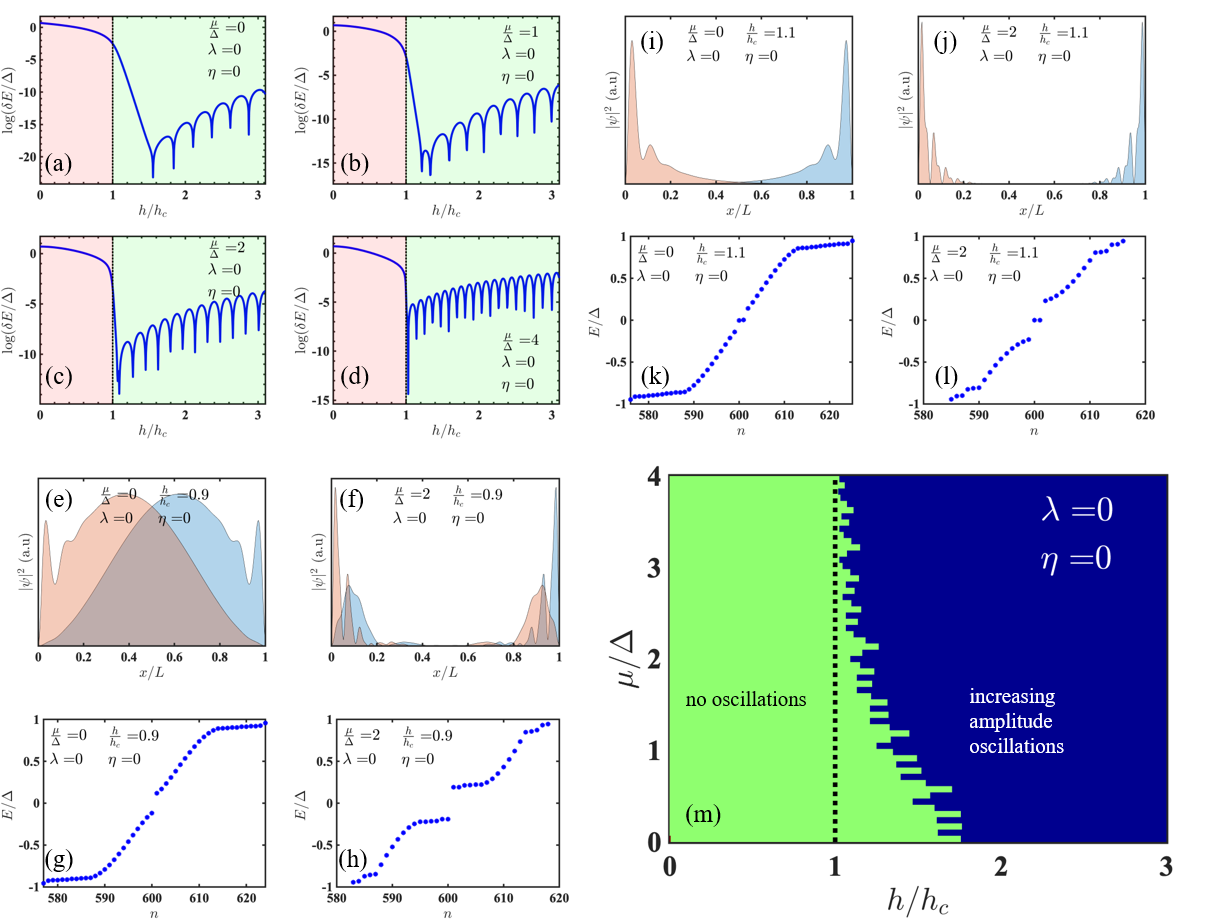}
    \caption{Numerical results for a pristine wire ($\lambda=\eta=0$, $V_{dot}=0$). (\textit{a}-\textit{d})  Energy splitting $\delta E$ as a function of $h/h_c$ for various values of the chemical potential $\mu$. The dashed vertical lines separate the trivial (pink) and topological (green) regions. (\textit{e-f}) Majorana components of the lowest energy states along with the corresponding  energy spectra (\textit{g-h}) in the trivial regime.  (\textit{i-j}) Majorana components of the lowest energy states and energy spectra (\textit{k-l}) in the topological regime. (m) ``Phase diagram'' showing regions characterized by no energy splitting oscillations (green) and oscillations that increase in magnitude with the magnetic field (blue).}    \label{Fig_lambda0_eta0}
    \vspace{2mm}
\end{figure*}

\section{Numerical results} \label{results} 

The numerical diagonalization of the (discretized) BdG Hamiltonian from Eq. (\ref{Eq_4}) gives us the energy spectrum and the wavefunctions corresponding to various parameter regimes. We focus on the lowest energy modes, which are either MBSs localized at the ends of the finite length wire (when $h>h_c$), or topologically-trivial ps-ABSs localized near the quantum dot  (when $h<h_c$). Note, however, that the non-local MBSs, as well as the component MBSs of the ps-ABSs, are always characterized by a finite overlap, which means that, strictly speaking, the Majorana modes are not (exact) eigenstates of the BdG Hamiltonian. In general, a pair of such partially overlapping MBSs corresponds to a finite energy fermionic (Bogoliubov) mode, which is what  the numerical solution actually describes. However, to gain physical insight, it is convenient to use the (formally-equivalent) Majorana representation and identify the MBS components of the lowest energy (fermionic) mode. Specifically, let us assume that  the lowest-energy solution of the BdG Hamiltonian is described by the spinor $\psi_{+\epsilon}(i) = (u_{i\uparrow},   u_{i\downarrow}, v_{i\uparrow},   v_{i\downarrow})^T$, where we use the basis  $(c_{i\uparrow}, c_{i\downarrow}, c_{i\uparrow}^\dagger, c_{i\downarrow}^\dagger)^T$.  Here, the index $i$ labels the sites of a 1D lattice, arrows indicate spin, $\epsilon$ is the energy of the eigenstate, while $u$ and $v$ designate the particle and hole components of the spinor, respectively. Due to particle-hole symmetry, there will also be a negative energy solution, $\psi_{-\epsilon}(i) = (v_{i\uparrow}^*,   v_{i\downarrow}^*, u_{i\uparrow}^*,   u_{i\downarrow}^*)^T$.
The component Majorana modes can be obtained from these solutions as the linear combinations
\begin{align}
\chi_A(i) &= \frac{1}{\sqrt{2}}\left[\psi_{\epsilon}(i) +\psi_{-\epsilon}(i)\right], \nonumber\\
\chi_B(i) &= \frac{i}{\sqrt{2}}\left[\psi_{\epsilon}(i) -\psi_{-\epsilon}(i)\right].     \label{Eq_chi}
\end{align}
The corresponding spinors in the Majorana basis can be written as  $\chi_\alpha(i)=(\widetilde{u}_{\alpha i\uparrow}, \widetilde{u}_{\alpha i\downarrow}, \widetilde{u}_{\alpha i\uparrow}^*, \widetilde{u}_{\alpha i\downarrow}^*)^T$, where $\alpha = A, B$ and $\widetilde{u}_{A,i,\sigma}=u_{i\sigma} +v_{i\sigma}^*$, while $\widetilde{u}_{B,i,\sigma}=i(u_{i\sigma} -v_{i\sigma}^*)$.
Note that the Majorana operator
\begin{eqnarray}
\gamma_\alpha^\dagger = \sum_i \left(\widetilde{u}_{\alpha i\uparrow}c_{i\uparrow}^\dagger + \widetilde{u}_{\alpha i\downarrow}c_{i\downarrow}^\dagger + \widetilde{u}_{\alpha i\uparrow}^*c_{i\uparrow} +  \widetilde{u}_{\alpha i\downarrow}^*c_{i\downarrow}\right),
\end{eqnarray}
manifestly satisfies the Majorana condition $\gamma_\alpha^\dagger = \gamma_\alpha$. Thus, for any pair of  BdG eigenstates  $\psi_{+\epsilon}$ and $\psi_{-\epsilon}$ one can uniquely define the corresponding Majorana modes $\chi_A$ and $\chi_B$. Using this construction, we study the properties of the low energy modes ($\epsilon\ll\Delta$) of the hybrid nanowire. Note that we have $\langle \chi_\alpha|H|\chi_\alpha\rangle=0$ and $\langle \chi_A|H|\chi_B\rangle=i\epsilon$, hence we can view the (finite) energy $\epsilon$ of the lowest energy mode as the energy splitting (or the hybridization energy) $\delta E \equiv \epsilon$ associated with the (finite) overlap of the component MBSs.

To characterize the overlap between the Majorana modes $\chi_A$ and $\chi_B$, let us define the characteristic length scale $\xi$ over which the Majorana components $\chi_\alpha$ decay exponentially and the distance $d_{AB}$ that characterizes  the separation between $\chi_A$ and $\chi_B$. 
We distinguish the following qualitatively different cases:  (i) $d_{AB}<\xi\ll L$, which corresponds to strongly overlapping Majorana modes forming a low energy  ``standard'' ABS, 
 (ii) $L>d_{AB}\gtrsim\xi$, which corresponds to partially separated MBSs forming a ps-ABS, and  (iii) $\xi\ll d_{AB}\sim L$, which corresponds to true, well separated Majorana modes localized at the ends of the wire.  We emphasize that the ps-ABSs, which occur in the topologically trivial phase,  i.e. for $h<h_c$, mimic the local phenomenology of (topological) Majorana zero modes as a result of local probes effectively coupling to only one of the (partially separated) Majorana components. One of the goals of this study is to identify possible qualitative differences between the energy splitting oscillations of ps-ABSs and those associated with topological MZMs. 

\vspace{2mm}
\subsection{Pristine wire} 
We start with a pristine, ``ideal'' system that does not contain a quantum dot or other inhomogeneity,  i.e. $\lambda=\eta=0$ and $V_{dot}=0$.
The results obtained by numerically diagonalizing the tight-binding BdG Hamiltonian are shown in Fig.~\ref{Fig_lambda0_eta0}. In the trivial regime ($h<h_c$), the energy $\delta E$ associated with the lowest energy mode  is on the order of the induced gap, decreases monotonically with the applied magnetic field and does not show any oscillatory features. By contrast, the topological regime ($h>h_c$) is characterized by energy splitting oscillations associated with the presence of (partially overlapping) MBSs at the ends of the finite system. Note that the amplitude of these oscillations increases with increasing magnetic field and increasing chemical potential. Also note that the   frequency of the oscillations  increases with the chemical potential. In addition, both quantities have a strong dependence on the length $L$ of the wire~\cite{sarma2012splitting} (not shown).  This behavior of the  splitting oscillations  is 
summarized in the ``phase diagram'' shown in Fig.~\ref{Fig_lambda0_eta0}(m), which is characterized by two distinct regions: (i) no oscillations (green area) and (ii) oscillations that increase in amplitude as a function of the magnetic field (blue).   The component Majorana wave functions shown in panels (e-f) and (i-j) reveal that the monotonic $\delta E$ behavior is associated with the presence of ABSs consisting of strongly overlapping MBS components, while the oscillatory behavior corresponds to the presence of well separated MBSs localized near the ends of the system and  having a finite (exponentially small) overlap. 

\begin{figure}[t]
    \centering
    \includegraphics[width=\columnwidth]{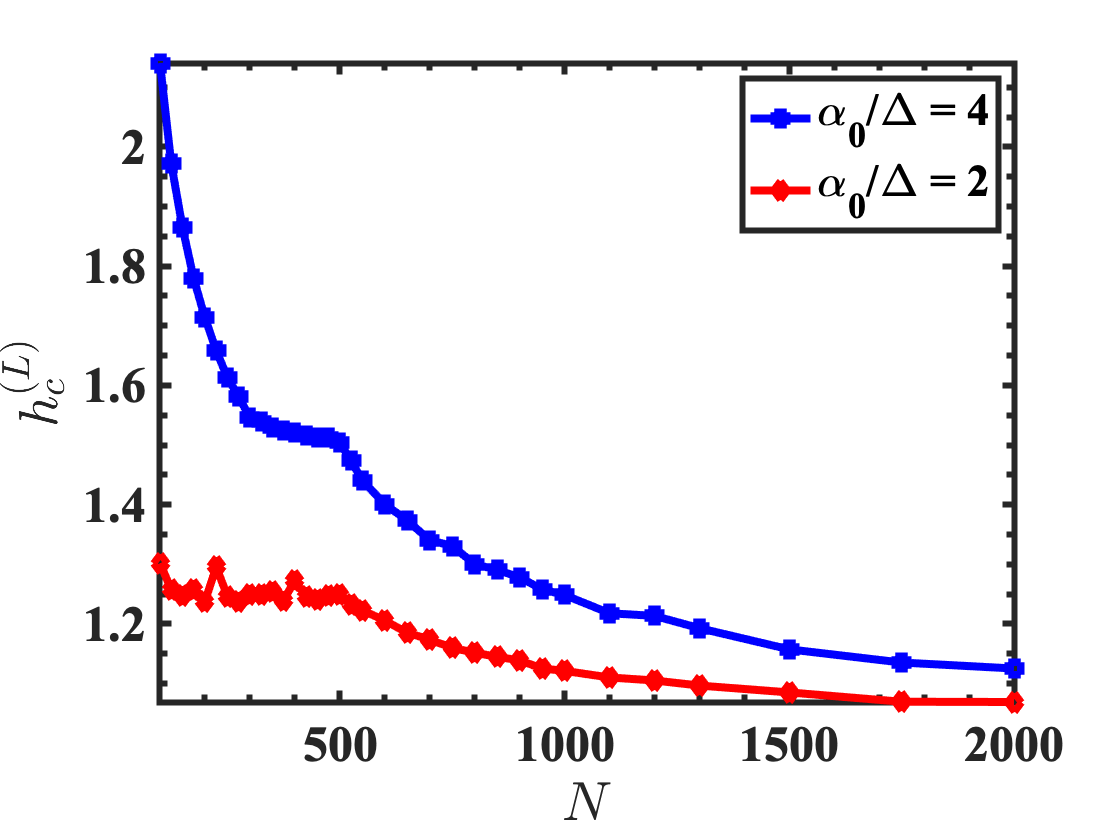}
    \caption{Dependence of the field $h_c^{(L)}$ associated with the first zero of the Majorana mode (in a pristine wire) on the size of the system (i.e., the number of lattice sites) for $\mu=0$ and two different values of the spin-orbit coupling. {Note that $h_c^{(L)}$ depends strongly on the spin-orbit coupling strength and approaches $h_c$ in the thermodynamic limit.}} \label{Fig_hcL}
    \vspace{-2mm}
\end{figure}

We note that the area characterized by the absence of oscillations (green region) extends into the nominally topological regime at low values of the chemical potential ($\mu \lesssim 2\Delta$). This is a finite size effect reflecting the fact that the first zero of the lowest energy mode (i.e., the Majorana mode) occurs at $h_c^{(L)}>h_c$, as one can clearly see in Fig. \ref{Fig_lambda0_eta0}(a). Note that the region $h_c < h < h_c^{(L)}$ is characterized by the presence of Majorana modes having long, non-oscillating exponential ``tails'' [see Fig. \ref{Fig_lambda0_eta0}(i)]  that become shorter with increasing $h$.  The overlap of these tails results in a finite hybridization energy that decreases monotonically with the field. 
The magnetic field $h_c^{(L)}>h_c$ that marks the onset of the oscillatory behavior depends on the chemical potential [as manifest in Fig. \ref{Fig_lambda0_eta0}(m)], the spin-orbit coupling strength, and the size of the system. More specifically, $h_c^{(L)}$ increases with increasing $\alpha$ and decreases slowly with the size $L$ of the system. This behavior is illustrated by the numerical results shown in Fig. \ref{Fig_hcL}. 
In the long wire limit, we have $h_c^{(\infty)}=h_c$ and the green (no oscillations) area coincides with the trivial phase ($h/h_c < 1$), while the blue (increasing oscillations) domain covers the entire topological phase  ($h/h_c > 1$). This property of the ``ideal'' model of the semiconductor-superconductor Majorana structure was the main reason behind the expected ``increasing amplitude'' behavior of the Majorana oscillations. This phenomenology, which holds for $h>h_c^{(L)}$ even in the presence of a quantum dot, as we explicitly show below, has not yet been observed experimentally. 

\begin{figure*}
    \centering
    \includegraphics[width=2\columnwidth]{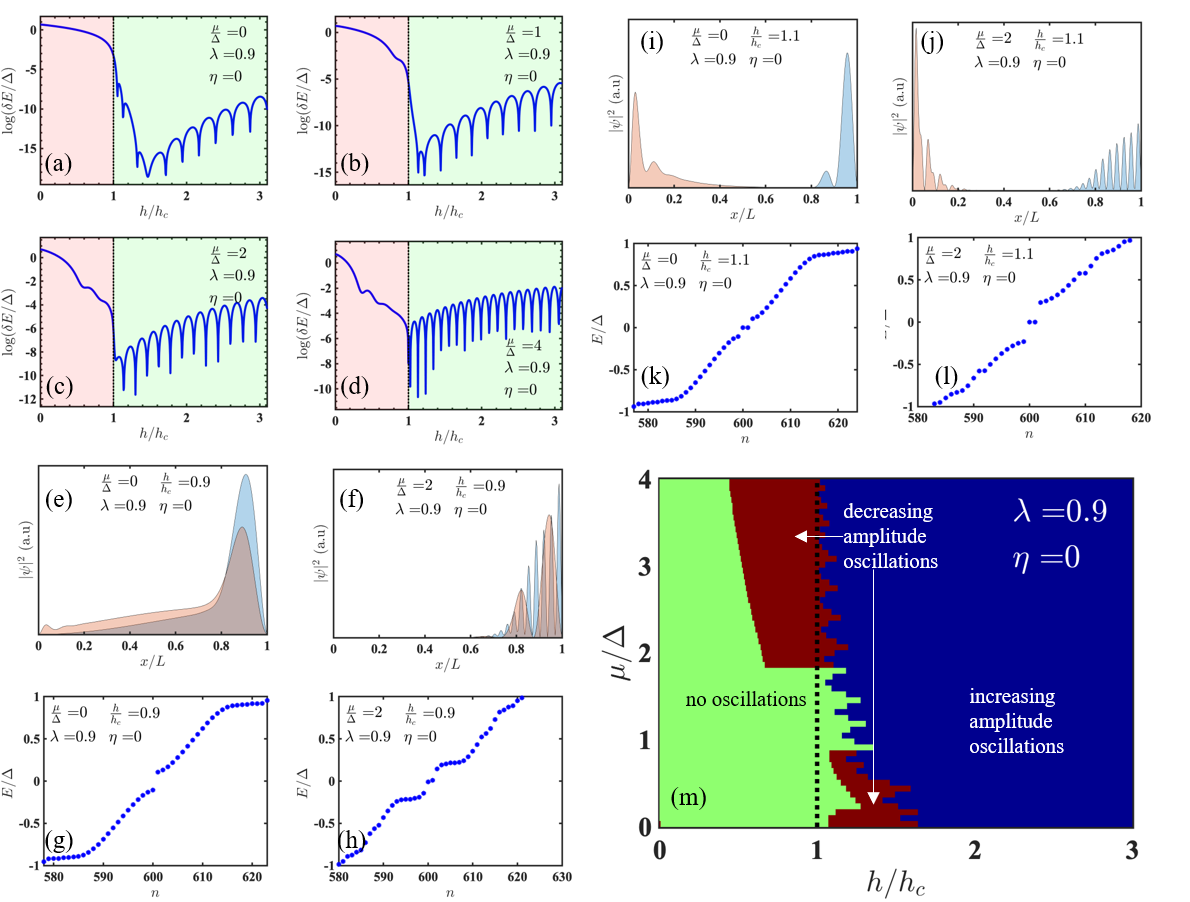}
    \caption{Numerical results for a system with step-like spin-orbit coupling ($\lambda=0.9$, $\eta=0$, $V_{dot}=0$). (\textit{a}-\textit{d})  Hybridization energy $\delta E$ as a function of $h/h_c$ for different values of the chemical potential $\mu$. The dashed vertical lines separate the trivial (pink) and topological (green) regions. (\textit{e-f}) Majorana components of the lowest energy states and the corresponding  energy spectra (\textit{g-h}) in the trivial regime.  (\textit{i-j}) Majorana components of the lowest energy states and energy spectra (\textit{k-l}) in the topological regime. (m) ``Phase diagram'' showing regions characterized by no oscillations (green) and oscillations that either decrease (dark red) or increase (blue) in magnitude with the magnetic field (blue). Note the small topological region characterized by decaying energy oscillations, as well as the (much larger) trivial decaying region corresponding to $\mu/\Delta \gtrsim2$.} \label{Fig_lambda0p9_eta0}
\end{figure*}

\subsection{Step-like spin-orbit coupling} 
Next, we consider a system with a position-dependent spin-orbit coupling (having a step-like profile), but with no quantum dot potential or suppressed pairing, i.e. $\lambda=0.9$, $\eta=0$, and $V_{dot}=0$. This type of model, which explicitly takes into account possible spatial variations of the spin-orbit coupling strength near the end of the system, was recently proposed in Ref. \onlinecite{cao2019decays} as a possible explanation for the  decaying oscillations observed experimentally. The results of our numerical calculations are shown in Fig.~\ref{Fig_lambda0p9_eta0} and are summarized by the ``phase diagram'' in panel (m). 
In the trivial regime, i.e., when $h<h_c$, $\delta E$ decreases almost monotonically with the applied field, but an oscillatory component develops when the chemical potential is larger than about $2\Delta$. The oscillatory component, which can be clearly seen in Figs. \ref{Fig_lambda0p9_eta0}(c) and \ref{Fig_lambda0p9_eta0}(d), is characterized by an amplitude that decreases with the magnetic field. Also,  Fig. \ref{Fig_lambda0p9_eta0} (f) reveals that this behavior is associated with an ABS consisting of two strongly overlapping MBSs localized at the right end of the system, i.e., in the region where the spin-orbit coupling is inhomogeneous. 
In the topological regime, $h>h_c$, the energy splitting $\delta E$ oscillates as a function of the magnetic field. Typically, i.e. for $h \gtrsim h_c^{(L-L^\prime)}$, where $h_c^{(L-L^\prime)}$ is defined by the first zero of the Majorana mode in a corresponding uniform wire (i.e. in the absence of the quantum dot), the magnitude of the oscillations increase with increasing magnetic field, similar to the ``ideal'' case discussed above. However, for small values of the chemical potential ($\mu \lesssim \Delta$) and magnetic fields in the vicinity of the critical value ($h_c < h \lesssim 1.4 h_c$), there is a small parameter widow characterized by oscillations of decreasing amplitude. Note that this window becomes very narrow above $\mu \eqsim \Delta/2$. 
The decaying oscillations can be clearly seen in Fig.~\ref{Fig_lambda0p9_eta0}(a). The Majorana nature of the underlying low-energy modes is revealed by the wave functions shown in Fig.~\ref{Fig_lambda0p9_eta0}(i). Note that the MBS localized at the right end of the system (i.e., in the region with reduced spin-orbit coupling) is characterized by a shorter localization length, as compared with its counterpart localized at the opposite end. Also note that in the trivial regime, the lowest energy mode is a non-degenerate ABS localized at the right end of the system  -- see  Fig. \ref{Fig_lambda0p9_eta0}(f) -- in contrast with the pristine wire case, where the lowest energy state is double degenerate, with ABSs localized at both ends of the system  -- see Fig. \ref{Fig_lambda0_eta0}(f). We conclude our discussion of this case by noting that decaying Majorana oscillations are possible, but observing them appears to require a high degree of fine tuning  [small dark red area in Fig. \ref{Fig_lambda0p9_eta0} (m)]. By contrast, increasing Majorana oscillations are generic [blue area in Fig. \ref{Fig_lambda0p9_eta0} (m)], even in the presence of inhomogeneous spin-orbit coupling. In addition, decaying oscillations appear to me more likely connected to topologically trivial states  [large dark red area in Fig. \ref{Fig_lambda0p9_eta0} (m)]. The robustness of these conclusions against possible variations of the relevant system parameters is tested below. 

\begin{figure*}
    \centering
    \includegraphics[width=2\columnwidth]{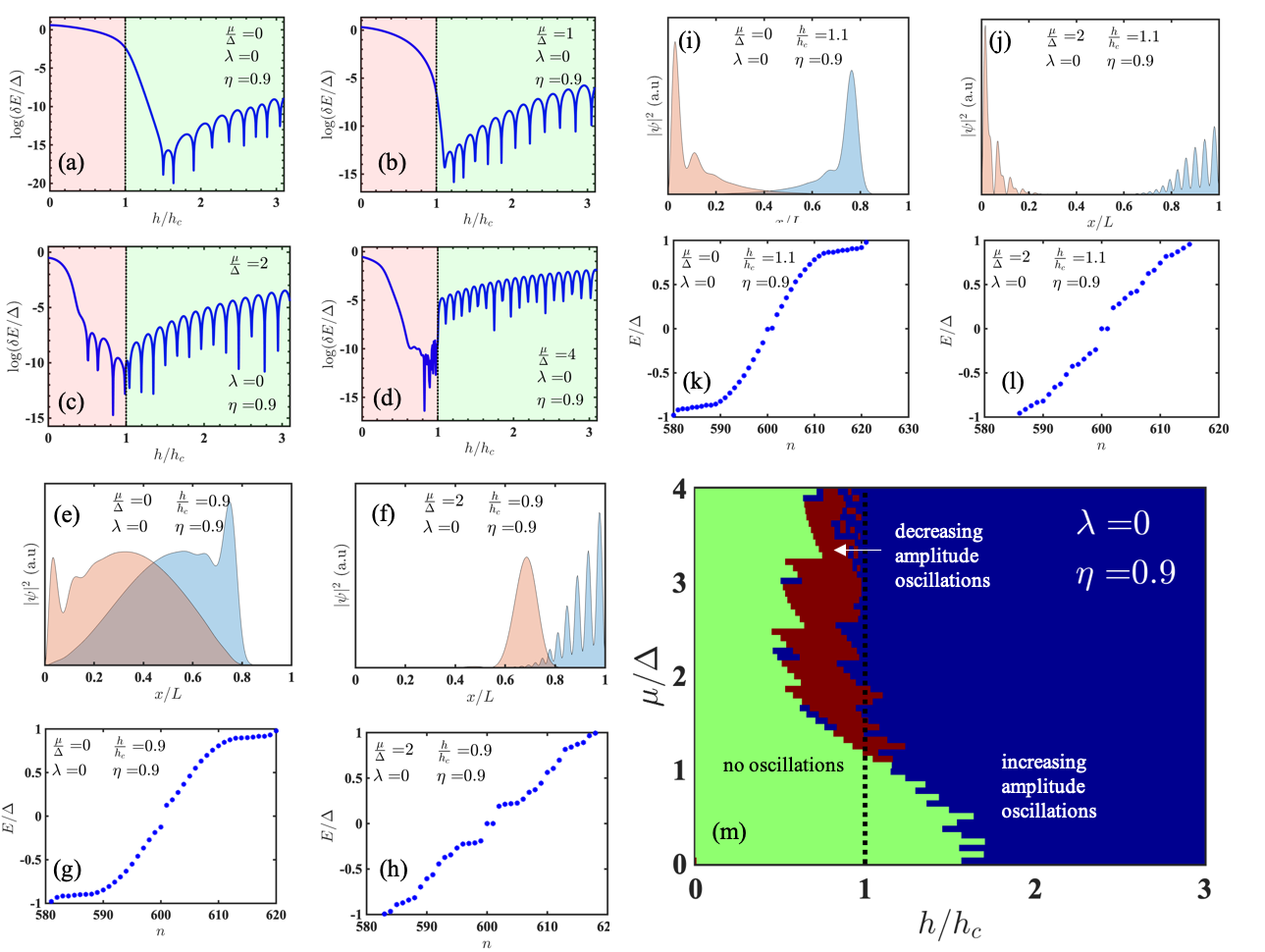}
    \caption{Numerical results for a quantum dot-wire system with uniform spin-orbit coupling ($\lambda=0$, $\eta=0.9$, $V_{dot}=2.7\Delta$). (\textit{a}-\textit{d}) Hybridization energy $\delta E$ as a function of $h/h_c$ for different values of the chemical potential $\mu$. (\textit{e-f}) 
Majorana components of the lowest energy states and energy spectra (\textit{g-h}) in the trivial regime.  (\textit{i-j}) Majorana components of the lowest energy states and energy spectra (\textit{k-l}) in the topological regime. (m) ``Phase diagram'' showing regions characterized by no oscillations (green) and oscillations that either decrease (dark red) or increase (blue) in magnitude with the magnetic field. Note that the region characterized by decaying energy oscillations is topologically trivial.
} \label{Fig_lambda0_eta0p9}
\end{figure*}

\subsection{Wire-dot system with uniform spin-orbit coupling} 
Consider now a wire coupled to a quantum dot. We assume that in the dot region the pairing potential is strongly suppressed ($\eta=0.9$) and that there exists a finite potential barrier of height $V_{dot}=2.7\Delta$, but we neglect (for now) the possible change in the spin-orbit coupling strength due to the presence of the dot ($\lambda=0$). The numerical results obtained by diagonalizing the tight-binding BdG Hamiltonian are shown in  Fig.~\ref{Fig_lambda0_eta0p9} and are summarized by the ``phase diagram'' in panel (m). 
For small values of the chemical potential ($\mu \lesssim \Delta$), $\delta E$ has a behavior that is very similar to that seen in pristine wires (see Fig. \ref{Fig_lambda0_eta0}), i.e., a monotonic decrease (with no oscillatory component) for $h < h_c^{(L-L^\prime)}$ and increasing amplitude oscillations for $h > h_c^{(L-L^\prime)}$. This behavior is due to the fact that all low-energy states are localized within the (homogeneous) proximitized wire region, as the potential barrier prevents them to penetrate into the quantum dot [see Figs. \ref{Fig_lambda0_eta0p9}(e) and \ref{Fig_lambda0_eta0p9}(i)]. 
\begin{figure*}
    \centering
    \includegraphics[width=2\columnwidth]{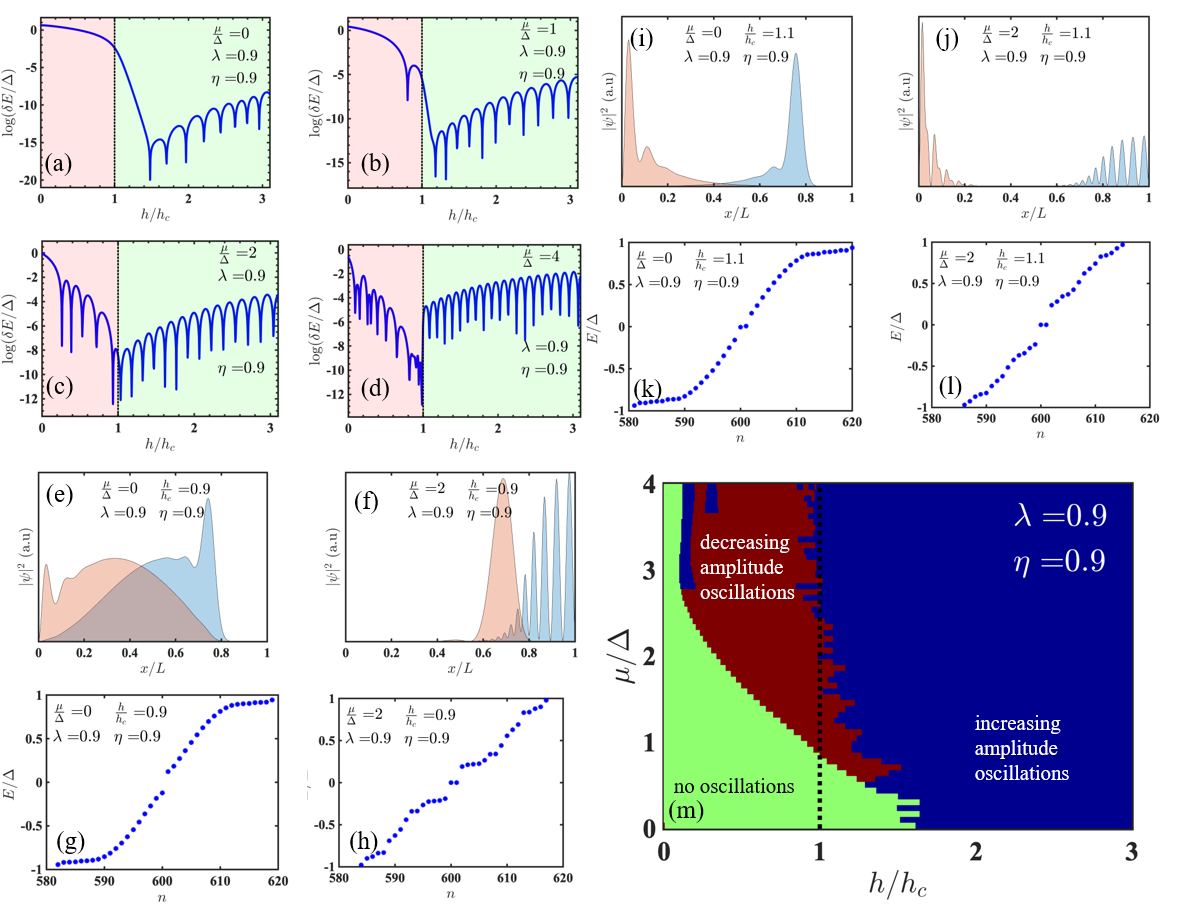}
    \caption{Numerical results for a quantum dot-wire system with position-dependent spin-orbit coupling, induced pairing, and effective potential ($\lambda=0.9$, $\eta=0.9$, $V_{dot}=2.7\Delta$).(\textit{a}-\textit{d}) Hybridization energy $\delta E$ as a function of $h/h_c$ for different values of the chemical potential $\mu$. (\textit{e-f})  Majorana components of the lowest energy states and energy spectra (\textit{g-h}) in the trivial regime.  (\textit{i-j}) Majorana components of the lowest energy states and energy spectra (\textit{k-l}) in the topological regime. (m) ``Phase diagram'' showing regions characterized by no oscillations (green) and oscillations that either decrease (dark red) or increase (blue) in magnitude with the magnetic field. Note that almost the entire region characterized by decaying energy oscillations is topologically trivial.} \label{Fig_lambda0p9_eta0p9}
\end{figure*}
Remarkably, for higher values of the chemical potential there is a significant topologically trivial region where $\delta E$ is characterized by oscillations that decrease in amplitude with the magnetic field  [dark red area in Fig.  \ref{Fig_lambda0_eta0p9}(m)]. The wave function of the Majorana components of a low-energy state responsible for this behavior is shown in Fig. \ref{Fig_lambda0_eta0p9}(f), revealing the fact that the decaying oscillations are associated with ps-ABSs consisting of fairly well separated MBSs localized near the quantum dot region. 
In the topological regime, the Majorana hybridization energy $\delta E$ oscillates with increasing amplitude, as as shown explicitly  in Fig.~\ref{Fig_lambda0_eta0p9}(a-d). Again, the onset of this oscillatory behavior corresponds to a field $h_c^{(L)} > h_c$ that depends on the chemical potential, the length of the system and the spin-orbit coupling strength (see Fig. \ref{Fig_hcL}). For $\mu>2\Delta$,  $h_c^{(L)}$ and $h_c$ practically coincide, while for $\mu\approx 0$ we have  $h_c^{(L)} < 2h_c$ for realistic system parameters.
Note that, in the presence of a potential barrier in the quantum dot region, there are practically no decaying oscillations in the topological regime, but there is a significant topologically trivial region where such decaying oscillations are present as a result of emerging near zero energy ps-ABSs. Also note that in certain conditions, e.g., for the parameters corresponding to Fig.  \ref{Fig_lambda0_eta0p9}(d), the typical energy of the ps-ABS is significantly lower that the energy of the topological Majorana mode that sets in for $h>h_c$.

\subsection{Wire-dot system: The general case} 
Finally, we consider the general case when the presence of the quantum dot induces inhomogeneity in all relevant parameters. Specifically we consider the case characterized by $\lambda=\eta=0.9$ and $V_{dot} = 2.7\Delta$. The corresponding numerical results are shown in Fig.~\ref{Fig_lambda0p9_eta0p9}. The ``phase diagram'' in panel (m) is qualitatively similar to the diagram in Fig.~\ref{Fig_lambda0_eta0p9}(m) discussed above. The only striking feature is that the topologically-trivial region characterized by ps-ABS-induced decaying oscillations is much larger in the presence of a reduced spin-orbit coupling in the quantum dot region, becoming the dominant low-field feature at finite chemical potential ($\mu\gtrsim 2\Delta$). Three observations are warranted. First, by comparing these results with those shown in Fig.~\ref{Fig_lambda0p9_eta0}, which correspond to the model used in Ref. \onlinecite{cao2019decays}, we notice that the small parameter window near $\mu=0$ and $h=h_c^{(L)}$ characterized by decaying Majorana oscillations vanishes in the presence of a finite effective potential. Since there is no strong reason to expect a position-dependent spin-orbit coupling in the presence of a uniform effective potential, we conclude that the likelihood of such parameter window actually existing in real systems is minimal. Second, in Fig. \ref{Fig_lambda0p9_eta0p9}(m) we observe that the dark red area associated with decreasing oscillations penetrates into the topological regime in a narrow region characterized by $h\approx h_c$ and  $\Delta \lesssim \mu\lesssim 2\Delta$. However, driving an actual hybrid system into this regime would require significant fine tuning. Moreover, if one is capable to observe this regime, one should also be capable to observe the nearby ``blue'' regime characterized by increasing Majorana oscillations. Third, the presence of a small topological region that is {\em not} characterized by increasing oscillations (i.e., the green and dark red areas with $h>h_c$) is a finite size effect that occurs for fields satisfying $h_c < h < h_c^{(L)}$. Since $h_c^{(L)} < 2h_c$ for realistic system parameters, the ability to access this region should imply the ability to access the nearby regime $h > h_c^{(L)}$ characterized by increasing Majorana oscillations. 

\section{Discussion} \label{discussion} 

So far,  we have examined the properties of the lowest energy mode  emerging in an effective model of a semiconductor-superconductor hybrid structure within different parameter regimes.  We have seen that the presence of a tunnel region at the end of the wire (which is the typical situation in a charge tunneling experiment) can generate  spin-orbit coupling inhomogeneity ($\lambda>0$) and a finite effective potential barrier ($V_{dot} >0$),  which, at finite chemical potential ($\mu \gtrsim \Delta$),  induce trivial low-energy Andreev bound states well before the topological quantum  phase transition corresponding to $h=h_c$.  In general, these low-energy states are ps-ABSs consisting of partially separated MBS components localized near the tunnel region at the end of the system, which can also be viewed as a quantum dot coupled to the proximitized wire. The presence of  low energy ps-ABSs can results in local signatures similar to those generated by topological  Majorana zero modes, e.g., quantized Majorana zero bias peaks in the differential conductance. This is due to the fact that only one of the partially separated MBS components couples measurably to a local probe (e.g., a normal lead) placed at the end of the system.  Further, the peak height can remain constant at $2e^2/h$ as a function of system parameters such as the magnetic field and tunnel barrier height resulting in quantized conductance plateaus~\cite{moore2018quantized}.

\begin{figure}
    \centering
    \includegraphics[width=\columnwidth]{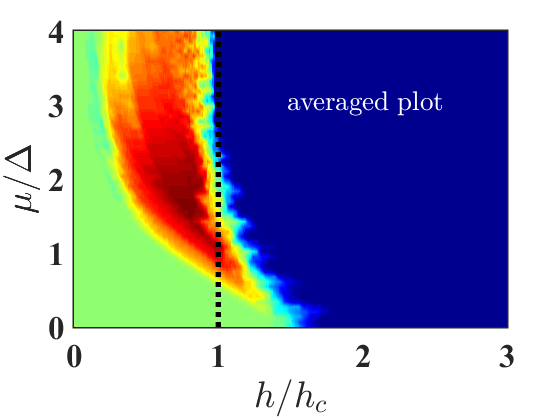}
    \caption{``Phase diagram'' averaged over different parameter configurations. Green, red, and blue correspond to energy splittings $\delta E$ 
characterized by no oscillations, decreasing amplitude oscillations, and increasing amplitude oscillations, respectively. The parameters used in the 
averaging procedure are combinations of  $\lambda\in\{0.1, 0.2, 0.3, 0.4\}$,  $\eta\in\{0.1, 0.2, 0.3, 0.4\}$, $V_{dot}/\Delta \in\{3, 4\}\eta$,
$\alpha/\Delta\in\{4,3\}$, and  $\beta_\Delta=\beta_\alpha =\beta_V\in \{0.05, 0.02\}$. We observe that almost the entire topological region ($h>h_c$) is characterized by increasing amplitude oscillations. By contrast, decreasing amplitude oscillations are likely to occur in the topologically trivial phase, i.e. at low values of the magnetic field. Accessing the narrow topological widow  $h_c < h < h_c^{(L)}$ characterized by deceasing Majorana oscillations or no oscillatory behavior requires fine tuning and places the system in close vicinity to the (generic) increasing Majorana oscillation regime (blue area).}
    \label{fig_avg}
\end{figure}

Since the observation of a robust zero bias conduction peak, or even of a quantized conductance plateau, cannot constitute a definitive demonstration of topological Majorana zero modes, we focus on the  hybridization energy $\delta E$ that  characterizes a pair of MBSs, as it contains non-local information associated with the overlap of the two Majorana modes. More specifically, we investigate the dependence of $\delta E$ on control parameters, such as the applied magnetic field and the chemical potential. 
In general, increasing the magnetic field affects both the location of the MBSs, as well as their localization length. In the topological regime, the two Majorana zero modes are already localized at the two ends of the nanowire. Therefore, increasing the magnetic field cannot enhance their spatial separation significantly, unless we consider very soft confinement (in which case the system will be definitely plagued with low-field ps-ABSs). 
However, the Majorana localization length increases with  the magnetic field and, therefore, the overlap of the two MZMs and the corresponding amplitude of the energy splitting oscillations also increase. By contrast, the ps-ABSs consist of partially separated MBSs localized on one side of the wire. Increasing the magnetic field significantly increases their separation, but has a much weaker effect on their localization length \cite{stanescu2019robust}. Consequently, increasing the magnetic field typically causes a suppression of the amplitude of ps-ABS-induced oscillations. This simple physical picture is  in general agreement with the numerical results presented above. In particular, decaying energy oscillations associated with topological Majorana zero modes are only present  within narrow parameter windows.  Driving the system within such a regime would require a significant degree of fine tuning and, most importantly, would put the system in the immediate vicinity of the (large) region characterized by increasing MZM-induced oscillations.  

In Sec.~\ref{results} we have presented results corresponding to several specific sets of parameters (spin-orbit coupling inhomogeneity, quantum dot potential, etc.). To demonstrate the robustness of our conclusions, we expand the parameter range by considering other possible parameter values. To simplify the presentation of the results and get a ``bird's view'' of the expected phenomenology, which should be realized in a typical experiment without a high degree of fine tuning, we average over  several (realistic) parameter configurations. Specifically, we consider all possible combinations involving 
 $\lambda\in\{0.1, 0.2, 0.3, 0.4\}$,  $\eta\in\{0.1, 0.2, 0.3, 0.4\}$, $V_{dot}/\Delta \in\{3, 4\}\eta$,
$\alpha/\Delta\in\{4,3\}$, and  $\beta_\Delta=\beta_\alpha =\beta_V\in \{0.05, 0.02\}$. The results are shown in Fig. \ref{fig_avg}. In the topological region ($h>h_c$), one generically expects Majorana-induced energy splitting oscillations of increasing amplitude (blue region), except a small window characterized by low-values of the chemical potential and magnetic fields immediately above $h_c$. On the other hand, the topologically trivial phase contains a large region characterized by (ps-ABS-induced) decaying energy oscillations (red area). Note that the low-field regime with no oscillations (green area) has no low-energy modes, i.e., $\Delta E$ is on the order of the induced gap.  The averaged ``phase diagram''  in Fig. \ref{fig_avg} reemphasizes the fact that, generically (i.e., without fine tuning the control parameters), the observation of decreasing amplitude oscillations at relatively low values of the magnetic field  is a clear signature of partially-separated Andreev bound states. By contrast, increasing amplitude of oscillations represent a strong signature of topological Majorana zero modes. 

\section{Conclusions} \label{conclusion} 

The prospect of fault tolerant topological quantum computation has inspired a large body of work to confirm the existence of Majorana bound states in 1D semiconductor-superconductor nanowire heterostructures. Given the far-reaching implications of this research, it is of paramount importance to carefully distinguish a true Majorana bound state from its doppelgangers, such as the partially-separated Andreev bound state. One approach that can be implemented experimentally within the current state-of-the-art is to carefully study of the hybridization energy $\delta E$ of the lowest energy modes as function of the applied magnetic field and the chemical potential. Recent observations~\cite{lin2012zero,cheng2009splitting,sarma2012splitting} indicate the presence of near-zero energy modes with energy splitting oscillations that decrease with increasing magnetic field, in sharp contrast with theoretical expectations based on simple modeling of the hybrid structure. It was suggested~\cite{cao2019decays} that an inhomogeneous spin-orbit coupling may be responsible for this behavior. In this work, we examine in detail the dependence of the hybridization energy on the applied magnetic field within a large parameter space that includes possible inhomogeneities near the ends of the wire, as expected for a typical charge conductance setup. We identify three main regimes: (i) the trivial phase, $h <h_c$, where $h_c=\sqrt{\Delta^2+\mu^2}$ is the critical field associated with the topological quantum phase transition, (ii) the oscillatory topological regime,  $h> h_c^{(L)}$, where $h_c^{(L)} \gtrsim h_c$ is associated with the first zero of the Majorana mode in an ``ideal'', uniform system of length $L$, and (iii) the anomalous topological regime $h_c < h < h_c^{(L)}$. Our extensive numerical results reveal that the oscillatory topological regime ($h> h_c^{(L)}$), which corresponds to almost the entire topological phase, is characterized by  increasing amplitude oscillations. So far, this behavior has not been observed experimentally. 

The anomalous topological regime represents a small parameter window characterized by $h_c < h < h_c^{(L)}$, with $h_c^{(L)}-h_c\sim \Delta$ for zero chemical potential and $h_c^{(L)}\approx h_c$ for $\mu \gtrsim 2\Delta$. Note that  $h_c^{(L)}=h_c$ in the limit of infinitely long wires, $L\rightarrow \infty$. In this regime the system could exhibit decaying Majorana oscillations or have no oscillatory behavior. However, accessing this regime 
requires a significant degree of fine tuning. Furthermore, once entering this regime, one should be able to observe the ``regular'', increasing amplitude Majorana oscillations at only slightly higher values of the magnetic field. 

Our results show that the trivial phase contains a significant region $h^* < h <h_c$ characterized by near-zero energy modes with decreasing amplitude oscillations. This feature is associated with the emergence of partially-separated Andreev bound states (ps-ABSs) -- also called quasi-Majorana modes\cite{vuik2018reproducing} -- that emerge generically in an inhomogeneous system at finite chemical potential. Remarkably, the characteristic field $h^*$ associated with the emergence of ps-ABSs can be much smaller than the critical field $h_c$, typically $h^* \ll h_c$  for $\mu > 2\Delta$.  Our analysis suggests that the low-energy features most likely to be observed at low values of the magnetic field in a semiconductor-superconductor structure are associated with near-zero energy ps-ABS. The corresponding energy splitting oscillations have amplitudes that decrease with increasing magnetic field. Since  the critical field $h_c$ can be much higher that the characteristic field $h^*$ associated with the emergence of ps-ABSs, the  topological regime (characterized by increasing amplitude Majorana oscillations) may be experimentally inaccessible. A direct test of these predictions (based on currently available devices) should involve studying the effect of local potentials (generated by gates placed near to ends of the system) on the energy splitting. Within the Majorana scenario, changes in the local potential at either end of the wire should affect the splitting, while a ps-ABS localized near one end of the system will be insensitive to change in the local potential near the opposite end.  

\begin{acknowledgments}
TDS was  supported  by  NSF DMR-1414683. CZ and ST acknowledge support from ARO Grant No. W911NF-16-1-0182. GS acknowledges IIT Mandi startup funds.
\end{acknowledgments}

\bibliography{biblio.bib}
\end{document}